\documentclass[twocolumn,showpacs,preprintnumbers,amsmath,amssymb]{revtex4}
\usepackage{graphicx}

 \newcommand\beq{\begin{equation}}
 
 \newcommand\eeq{\end{equation}}
 \newcommand\beqn{\begin{eqnarray}}
 \newcommand\eeqn{\end{eqnarray}}
 \newcommand\GeV{{\rm GeV}}
 
\def\fm{\,\mbox{fm}}
\def\GeV{\,\mbox{GeV}}

\def\lsim{\mathrel{\rlap{\lower4pt\hbox{\hskip1pt$\sim$}}
    \raise1pt\hbox{$<$}}}         
\def\gsim{\mathrel{\rlap{\lower4pt\hbox{\hskip1pt$\sim$}}
    \raise1pt\hbox{$>$}}}         
\def\BA{\begin{eqnarray}}
\def\BE{\begin{equation}}
\def\BF{\begin{figure}[htb]}
\def\BT{\begin{table}[htb]}
\def\EA{\end{eqnarray}}
\def\EE{\end{equation}}
\def\EF{\end{figure}}
\def\ET{\end{table}}

\begin{document}


\title{
Study of Nuclear Suppression
at Large Forward Rapidities in d-Au Collisions
at Relativistic and Ultrarelativistic Energies
}

\author{J.~Nemchik$^{a,b}$}
\author{V.~Petr\' a\v cek$^b$}
\author{I.K.~Potashnikova$^{c,d}$}
\author{M.~\v Sumbera$^e$}

\affiliation{
$^a$
Institute of Experimental Physics SAS, Watsonova 47,
04001 Ko\v sice, Slovakia
\\
$^b$
Czech Technical University,
FNSPE, Brehova 7,
11519 Praque, Czech Republic
\\
$^c$
Departamento de F\'{\i}sica y Centro de Estudios Subat\'omicos, 
Universidad T\'ecnica Federico Santa Mar\'{\i}a,
Casilla 110-V, Valpara\'{\i}so, Chile
\\
$^d$
Joint Institute for Nuclear Research, Dubna, Russia
\\
$^e$
Nuclear Physics Institute AS CR,
25068 \v Re\v z/Prague, Czech Republic
}

\date{\today}

\begin{abstract}

We study a strong suppression of the relative
production rate $(d-Au)/(p-p)$ for inclusive high-$p_T$
hadrons of different species at large forward rapidities
(large Feynman $x_F$).
The model predictions calculated in the light-cone
dipole approach are in a good agreement with the recent
measurements by the BRAHMS and STAR Collaborations
at the BNL Relativistic Heavy Ion Collider.
We predict a similar suppression at large $p_T$
and large $x_F$ also at lower energies, where no effect
of coherence is possible. It allows to exclude the saturation
models or the models based on Color Glass Condensate
from interpretation of nuclear effects.

\end{abstract}


\maketitle

%
%
\section{Introduction}
\label{intro}
%
%

In the proton(deuteron)-nucleus collisions,
investigated at the Relativistic Heavy Ion
Collider (RHIC),
recent measurements of high-$p_T$ particle spectra
at large forward rapidities
performed recently by the BRAHMS \cite{brahms,brahms-07}
and STAR \cite{star} Collaborations show
a strong nuclear suppression.
The basic explanation for such an effect has been based
on an idea that in this kinematic region 
corresponding to the beam fragmentation region
at large Feynman $x_F$ one can reach the smallest values
of the light-front momentum fraction variable $x_2$ in nuclei.
It allows to access the strongest coherence effects
such as those associated with shadowing or the Color Glass
Condensate (CGC).

It was shown in \cite{knpsj-05,knpsj-05c} that a considerable nuclear
suppression for any reaction at large $x_F$ (small $x_2$)
is caused by another effects, which can be easily
misinterpreted as coherence. 
Such a suppression, for example, 
can be treated in terms of the Sudakov form factor reflecting
the energy conservation. 
It is governed by the probability to produce no particles
at large $x_F\to 1$.

Nuclear suppression at large $x_F$ can be also 
interpreted alternatively, as a consequence of a reduced
survival probability for large rapidity gap (LRG) processes
in nuclei, an enhanced resolution of higher Fock 
states by nuclei, or an effective energy loss that
rises linearly with energy.
It was demonstrated in refs.~\cite{knpsj-05,knpsj-05c} that it 
is a leading twist effect, violating QCD factorization.

The BRAHMS Collaboration \cite{brahms} in 2004 for the first time
found a substantial nuclear suppression for high-$p_T$
negative hadrons produced at large pseudorapidity $\eta = 3.2$
(see Fig.~\ref{brahms}).
Because the data cover rather small $x_2\sim 10^{-3}$, the
interpretation of such a suppression has been tempted to be
as a result of saturation \cite{glr,al} or the
CGC \cite{mv}, expected in some models \cite{kkt}.

Alternative interpretation of the nuclear effects
occurring in the BRAHMS data \cite{brahms}
is based on energy conservation
implemented into multiple soft rescatterings
of the projectile quark in nuclear matter
~\cite{knpsj-05,knpsj-05c}.
Moreover, new data for neutral pions from the STAR Collaboration
have been recently appeared at the same c.m. energy, $\sqrt{s} = 200\,$GeV
but at still larger pseudorapidity $\eta = 4.0$ demonstrating
a huge nuclear suppression, which is more than a factor of 2 larger
than at $\eta = 3.2$ manifested by the BRAHMS data. 
Although a minor part of this difference can be explained by the isospin
effects, i.e. by a difference in production of $h^-$ and $\pi^0$
particles on deuteron target, there is still a large room for investigation
of this huge suppression. 
It provides a good possibility to test our
interpretation of such an effect. 
This represents one of the main goal
of the present paper. 

Another very interesting effect 
following from our interpretation
of nuclear effects at large $x_F$ and
supported by 
available data is the $x_F$ scaling of nuclear
suppression. It is in contradiction
with $x_2$ scaling, which is expected if the scaling represents
the net effect of quantum coherence.
The detailed analysis of this $x_F$ scaling can be found
in \cite{knpsj-05,knpsj-05c} collecting the data for
production of different species of leading hadrons with small $p_T$ 
in $p-A$ collisions at different energies
covering the laboratory energy range from 70 GeV to 400 GeV.
Similar $x_F$ scaling of the nuclear effects is observed 
from data on $J/\Psi$ and $\Psi'$ production measured
by the E866 Collaboration at Fermilab \cite{e866-psi}
compared to lower-energy data \cite{na3}. Besides,
recent measurements of nuclear suppression for $J/\Psi$
production in $d-Au$ collisions by the PHENIX Collaboration
\cite{psi-qm} at RHIC are consistent with $x_F$ scaling
and exhibit a dramatic violation of $x_2$ scaling when
compared with the E866 data \cite{e866-psi}.

According to $x_F$ scaling
\cite{knpsj-05,knpsj-05c},
we expect similar nuclear effects at forward rapidities 
also at lower energies when the onset of coherence effects
is much weaker. It gives much less room for explanation
of a strong nuclear suppression in terms of CGC.
Because new data of high-$p_T$ hadron production 
at forward rapidities are expected also at 
smaller c.m. energies $\sqrt{s} = 130\,$ GeV
and $62.4\,$GeV, corresponding predictions for nuclear effects
at large $p_T$ will play an important role 
for further verification of various
phenomenological models based on the CGC.

Besides, new data from the BRAHMS Collaboration \cite{brahms-07}
for different species of high-$p_T$ hadrons 
produced at $\eta = 3.0$ have been recently appeared.
It allows to provide another probe of our model in investigation
of significant nuclear suppression at forward rapidities.
This coincides with the further goal of the present paper. 

The paper is organized as follows. 
In Sect.~\ref{scatt} we present shortly
a formulation of nuclear suppression in terms
of Sudakov suppression factors adopted
for multiple parton interactions with the nucleus.
Here we also mention about
three different mechanisms of high-$p_T$ particle
production.
In the next Sect.~\ref{data} we
calculate the predictions for the ratio of particle production rate
in $d-A$ and $p-p$ collisions as a function of $p_T$ at large
forward rapidities corresponding to BRAHMS and STAR experiments
at RHIC. We analyze much stronger nuclear suppression
obtained recently by the STAR Collaboration \cite{star} at
$\eta = 4.0$ in comparison with the well known results
from the BRAHMS experiment \cite{brahms} at
$\eta = 3.2$. We demonstrate that the
model calculations without any free
parameter are in a good agreement with data from the both
collaborations.
Finally we perform predictions for nuclear
suppression at large forward rapidities also at lower
energies, where no effect of coherence is possible.
It allows to exclude the models based on CGC
from explanation of a strong nuclear suppression.
We also demonstrate an approximate  
$\exp(\eta)/\sqrt{s}$ ($x_F$) scaling
of nuclear effects in the energy and pseudorapidity range
accessible by the BRAHMS and STAR Collaborations.
The results of the paper are
summarized and discussed in Sect.~\ref{conclusions}.

%
%
\section{High-$p_T$ hadron production at forward rapidities:
Sudakov suppression, production cross section}
\label{scatt}
%
%

Any hard reaction in the limit $x_F\to 1$ can be treated as
LRG process, where
gluon radiation is forbidden by energy conservation.
If a large-$x_F$ particle is produced, the rapidity interval to be kept
empty is $\Delta y=-\ln(1-x_F)$. 
Assuming as usual an uncorrelated Poisson distribution for
radiated
gluons, the Sudakov suppression factor, i.e. the probability to have a
rapidity gap $\Delta y$, was developed in ref.~\cite{knpsj-05}
and has a very simple form,
%
%
 \beq
S(x_F)=1-x_F\ .
\label{60}
 \eeq
%
%

Nuclear suppression at $x_F\to 1$ can be formulated as a survival
probability of the LRG in multiple interactions with the nucleus.
Every additional inelastic interaction contributes an extra suppression
factor $S(x_F)$. The probability of an n-fold inelastic collision is
related to the Glauber model coefficients via the
Abramovsky-Gribov-Kancheli (AGK)  cutting rules \cite{agk}.
Correspondingly, the survival probability at impact parameter $\vec b$
reads,
%
%
 \beqn
W^{hA}_{LRG}(b) &=&
\exp[-\sigma_{in}^{hN}\,T_A(b)]
\nonumber\\ &\times&
\sum\limits_{n=1}^A\frac{1}{n!}\,
\left[\sigma_{in}^{hN}\,T_A(b)\,
\right]^n\,S(x_F)^{n-1}\ .
 \label{70}
 \eeqn
%
%
where $T_A(b)$ is the nuclear thickness function.

Assuming
large values of hadron transverse momenta,
the cross section of hadron production in $dA(pp)$ collisions is given by
a convolution of the distribution function for the projectile valence
quark with the quark scattering cross section and the fragmentation
function,
%
%
 \beqn
&&\frac{d^2\sigma}{d^2p_T\,d\eta} =
\sum\limits_q \int\limits_{z_{min}}^1 dz\,
f_{q/d(p)}(x_1,q_T^2)
\nonumber\\ &\times&
\left.\frac{d^2\sigma[qA(p)]}{d^2q_T\,d\eta}
\right|_{\vec q_T=\vec p_T/z}\,
D_{h/q}(z),
\label{80}
 \eeqn
%
%
where
$x_1=q_T\,e^\eta\,/\sqrt{s}$.
The quark distribution functions in the nucleon have the form using
the lowest order parametrization of Gluck, Reya and Vogt \cite{grv}. 
For fragmentation functions we use parametrization from \cite{fs-07}.

The main source of suppression at large $p_T$ concerns
to multiple quark rescatterings
in nuclear matter.
The quark distribution in the nucleus
can be defined performing summation
over multiple interactions and 
using the probability of an $n$-fold inelastic collision
related to the Glauber model coefficients with Gribov's
corrections via AGK cutting rules \cite{agk}.
It has the following form:
%
%
\BE
f_{q/N}^A(x,q_T^2,\vec{b},z) =
\sum\limits_{n=0}^{A}\,v_n(\vec{b},z)\,f_{q/N}^{n}(x,q_T^2)\, ,
\label{100}
\EE
%
%
where the coefficients $v_n$ reads
%
%
\BE
v_n(\vec{b},z) =
\frac{\biggl [\sigma_{eff}\,T(b,z)\biggr ]^n}
{\biggl [1 + \sigma_{eff}\,T(b,z)\biggr ]^{n+1}}\, .
\label{110}
\EE
%
%
The effective cross section
$\sigma_{eff}$
was evaluated in \cite{knpsj-05}.

The valence quark distribution functions $f_{q/N}^{n}(x,q_T^2)$
in Eq.~(\ref{100}) are also given by the GRV parametrization
\cite{grv} but contain extra suppression factors, $S(x)^n = (1 - x)^n$
(\ref{60}), corresponding to an $n$-fold inelastic collision,
%
%
\BE
f_{q/N}^{n}(x,q_T^2)
=
C_n\,f_{q/N}(x,q_T^2)\,S(x)^n
\, ,
\label{120}
\EE
%
%
where the normalization factors $C_n$ are fixed by the Gottfried
sum rule.

The cross section of quark scattering on the target
$d\sigma[qA(p)]/d^2q_Td\eta$
in Eq.~(\ref{80}) is
calculated in the light-cone dipole approach \cite{zkl,jkt-01}.
Performing calculations, 
we separate the contributions characterized by different
initial transverse momenta of the projectile partons
and sum over different mechanisms of
high-$p_T$ production.
 
{\bf Quark-diquark break up of the proton.}
Here we consider proton breakup
remaining the diquark intact, $p\to \widehat
{qq}+q$.
We treat the diquark $\{qq\}$ as point-like and integrate over
its momentum.
The corresponding $k_T$ distribution
of the projectile valence quark,
after propagation through the nucleus at impact parameter $\vec b$, is   
is calculated using the dipole technique developed
in refs.~\cite{kst1,kst2} (see also \cite{knpsj-05}).  
This contribution dominates the low transverse momentum region $k_T \lsim
1\GeV$.

{\bf Diquark break up \boldmath$\widehat{qq}\to qq$.}  
At larger $k_T$ the interaction resolves the diquark, so its break-up
should be included.
In this case the valence quark has much larger primordial
transverse momentum. Its contribution is calculated
in accordance with \cite{kst1,kst2} (see also \cite{knpsj-05}).

{\bf Hard gluon radiation \boldmath$q\to Gq$.}
At large $k_T$ the dipole approach should recover the parton model
\cite{3f},
where high momentum transfer processes occur (in leading order)
as binary collisions with the transverse momentum of each final parton of
order $k_T$.
Therefore, one
should explicitly include in the dipole description radiation of a gluon
with large transverse momentum that approximately equilibrates $k_T$, 
i.e. the process $qN\to qGX$.
We employ the nonperturbative quark-gluon wave function developed in
\cite{kst2}, which corresponds to small gluonic spots in the nucleon
\cite{kp,kpp-07}. Details of calculation of the quark scattering
cross section can be found in \cite{knpsj-05}.

%
%
\section{Comparison with data}
\label{data}
%
%

\begin{figure}[tbh]
\includegraphics{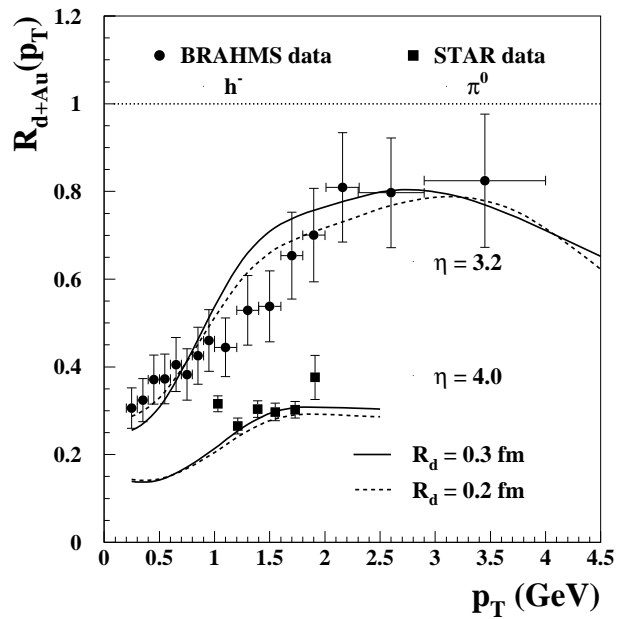}
\vspace{8cm}
{\caption[Delta]
 {Ratio, $R_{d+Au}(p_T)$, of negative particle production rates in $d-Au$ and 
$p-p$
collisions as function of $p_T$ at pseudorapidity $\eta = 3.2$ vs. data from
the BRAHMS Collaboration \cite{brahms}. Ratio $R_{d+Au}(p_T)$ for neutral pion
production at $\eta = 4.0$ vs. data from the STAR Collaboration \cite{star}.
Solid and dashed curves correspond to calculations with the diquark size $0.3\fm$
and $0.2\fm$ respectively.  
}
 \label{brahms}}
 \end{figure}

Several years ago 
the BRAHMS collaboration performed measurements of 
nuclear effects at pseudorapidity $\eta=3.2$ for
production of negative hadrons with transverse
momentum up to $p_T\approx 3.5\GeV$. Instead of the usual Cronin enhancement
a suppression was found, as one can see from Fig.~\ref{brahms}.

Rather strong nuclear suppression of data at small $p_T$
has been analysed and interpreted in details
in refs.~\cite{knpsj-05,knpsj-05c} and such an analysis
does not need to be repeated here.
On the other hand
we will concentrate on a study of nuclear effects
at large $p_T$.

Note that the dominance of valence quarks in the projectile proton leads
to an isospin-biased ratio. Namely, high-$p_T$ negative hadrons 
close to the kinematic limit are produced mainly from $d$, rather than
$u$, quarks.  Therefore, more negative hadrons are produced by deuterons
than by protons, and this causes an enhancement of the ratio plotted in
Fig.~\ref{brahms} by a factor of $3/2$. 
We take
care of this by using proper fragmentation functions from \cite{fs-07}.

Although the nuclear effects under discussion are not
sensitive to $p_T$ dependence
of the cross section for hadron production in $p-p$
collisions, the
model has been already successfully confronted with
$p-p$ data from the BRAHMS
experiment \cite{brahms} 
at $\eta=3.2$ in refs.~\cite{knpsj-05,knpsj-05c}. 

As the next step,
very important for verification of our model, we
calculate nuclear effects employing the dipole
formalism and the mechanisms described shortly in the previous section
(see also~\cite{knpsj-05}).
The results are compared
with the BRAHMS data for the minimum-bias ratio $R_{d+Au}(p_T)$
\cite{brahms} in Fig.~\ref{brahms}. 
One can see that calculations are in a rather good
agreement with data.

In the same Fig.~\ref{brahms} we show also the STAR data 
for $\pi^0$ production
presented as the ratio $R_{d+Au}(p_T)$ at pseudorapidity $\eta = 4.0$
\cite{star}.
Data demonstrate that suppression is much stronger than at $\eta = 3.2$
observed by the BRAHMS Collaboration \cite{brahms}. 
A part of this difference can be explained by an isospin-biased
ratio followed from the dominance of valence quarks in the projectile
proton (deuteron). Whereas more negative hadrons are produced by
deuterons than by protons, for positive hadrons the situation is opposite.

 \begin{figure}[tbh]
\includegraphics{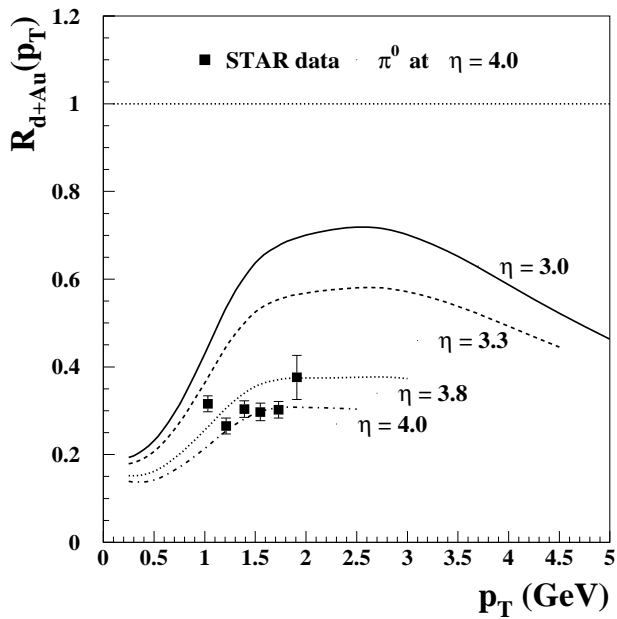}
\vspace{8cm}
{\caption[Delta]
 {
Model predictions for 
nuclear attenuation factor $R_{d+Au}(p_T)$ as a function of
transverse momentum for production of $\pi^0$ mesons
at $\sqrt{s}=200\GeV$ and
different values of pseudorapidity $\eta$ from $3.0$ to $4.0$. 
For an illustration, full squares represent the data from
the STAR Collaboration \cite{star} obtained at $\eta = 4.0$.}
 \label{eta-pi0}}
 \end{figure}

For production of negative hadrons, it leads to an enhancement
of the ratio $R_{d+Au}(p_T)$ by a factor of $3/2$ in comparison
with $R_{p+Au}(p_T)$. However, for production of positive
hadrons one arrives to a suppression by a factor of $3/4$.
As a result, for $\pi^0$ mesons one gets 
for $R_{d+Au}(p_T)$ a small overall suppression factor $= 5/6$,
which is smaller than a factor of $3/2$ for negative hadrons.
However, such a difference following from the isospin effects can explain
only a minor part from a huge difference in nuclear
suppressions experimentally observed at $\eta = 3.2$ and
$4.0$ by the BRAHMS \cite{brahms} and STAR \cite{star}
Collaborations, respectively.

If one supposes to interpret the nuclear effects
of high-$p_T$ hadron production at $\eta = 3.2$
in terms of CGC, such an interpretation should fail
at larger $\eta = 4.0$, where
observed suppression by the STAR Collaboration \cite{star}
is more than a factor of 2 larger
than at $\eta = 3.2$ manifested by the BRAHMS data \cite{brahms}.
The stronger onset of the quantum coherence effects at
$\eta = 4.0$ in comparison with $\eta = 3.2$ can not
explain such a huge rise of nuclear suppression.

Much stronger onset of nuclear effects
at $\eta = 4$
(at larger Feynman $x_F$) can be simply explained only
by energy conservation and
reflects a much smaller
survival probability of LRG in multiple quark interactions
at larger pseudorapidity \cite{knpsj-05,knpsj-05c}.

As a demonstration 
of different onsets of nuclear effects
as a function of pseudorapidity
we present in Fig.~\ref{eta-pi0} predictions
for
nuclear suppression at different fixed
values of $\eta$ calculating
nuclear attenuation factor $R_{d+Au}(p_T)$ 
for production of $\pi^0$ mesons
at $\sqrt{s}=200\GeV$.
Changing the value of $\eta$ from $3.0$ to $4.0$,
one can see a large rise of nuclear suppression
about a factor of $2$.

 \begin{figure}[tbh]
\includegraphics{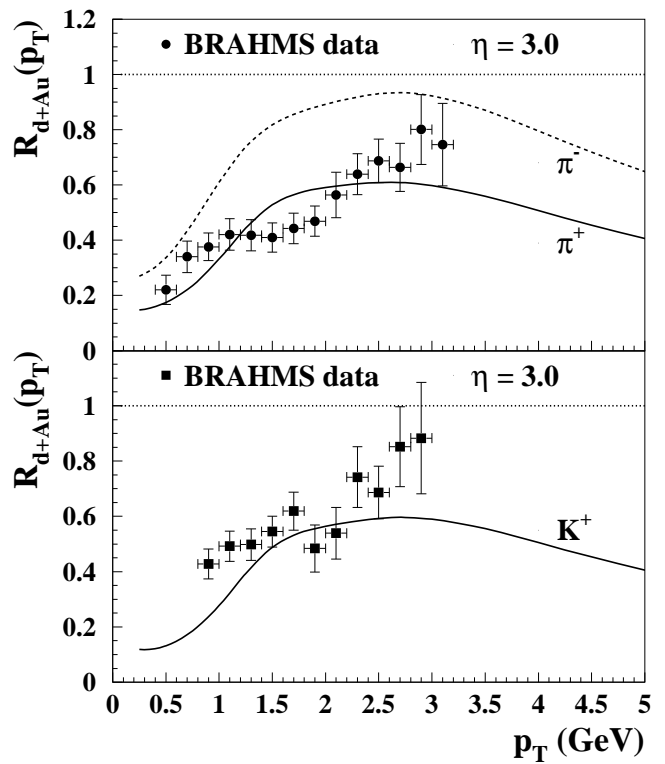}
\vspace{10cm}
{\caption[Delta]
{Ratio, $R_{d+Au}(p_T)$, of identified particle production rates in $d-Au$ and 
$p-p$ collisions as function of $p_T$ at pseudorapidity $\eta = 3.0$ vs. data 
from the BRAHMS Collaboration \cite{brahms-07}. 
}
 \label{brahms-07}}
 \end{figure}

The BRAHMS Collaboration has recently reported 
new data \cite{brahms-07} measuring nuclear effects
for production of different species of hadrons 
at $\eta = 3.0$ in $d+Au$ collisions.
These new data confirm a substantial nuclear suppression,
which is similar to what has been found already in 2004
\cite{brahms}.
Using proper fragmentation functions
from \cite{fs-07},
we present in Fig.~\ref{brahms-07} the model
predictions together with the last BRAHMS data 
at $\eta = 3.0$ for production of $\pi^+$
and $K^+$ mesons. One can see again a reasonable
agreement of the model with data.

As a demonstration of
the valence quark domination in the projectile particle
leading to an enhancement in production of negative hadrons by deuterons,
we present in the same
Fig.~\ref{brahms-07} also the model predictions for $\pi^-$ production
at $\eta = 3.0$.
Much smaller nuclear suppression clearly confirms
the isospin asymmetry of leading particle production
at large forward rapidities and large $p_T$
in $d-Au$ collisions.

 \begin{figure}[tbh]
\includegraphics{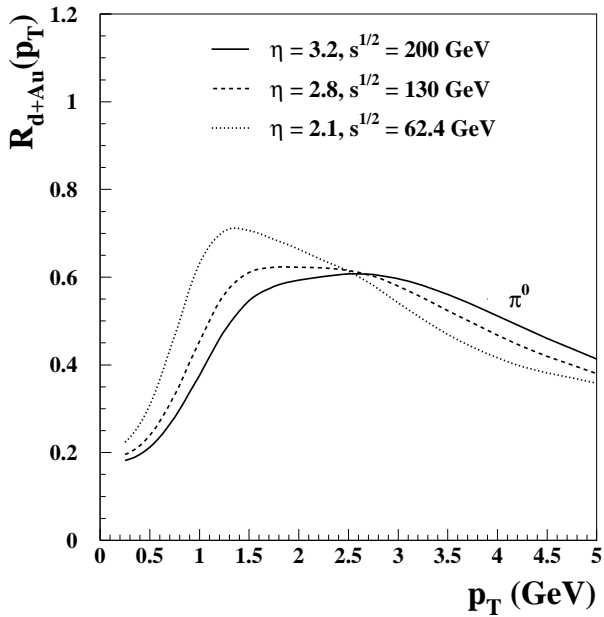}
\vspace{8cm}
{\caption[Delta]
{Theoretical predictions for an approximate $\exp(\eta)/\sqrt{s}$ scaling
of the nuclear attenuation factor $R_{d+Au}(p_T)$
for neutral pion production in the energy and pseudorapidity range
accessible by the BRAHMS and STAR Collaborations.
}
 \label{scaling}}
 \end{figure}

Energy conservation applied to multiple parton
rescatterings in nuclear medium leads to
$x_F$ scaling of nuclear effects
\cite{knpsj-05,knpsj-05c}.
We expect approximately the same
nuclear effects at different energies and pseudorapidities
corresponding to the same values of $x_F$.
Such a situation is demonstrated in Fig.~\ref{scaling},
where we present $p_T$ dependence of nuclear attenuation
factor $R_{d+Au}(p_T)$ for $\pi^0$ mesons
at different c.m. energies $\sqrt{s} = 200, 130$ and  
$62.4\,$GeV and such corresponding values of $\eta$, which
keep the same value of $x_F$.
Such a $x_F$ scaling can be verified and investigated
in the future by the
BRAHMS or STAR Collaborations in the energy and pseudorapidity
range accessible at RHIC also for production of other
species of hadrons.

%
%
%
\section{Summary and conclusions}\label{conclusions}
%
%

In the present paper we analyze a significant nuclear suppression
in production of different species of particles
at large pseudorapidities (large $x_F$) in $d-Au$ collisions investigated
at present mainly by the BRAHMS \cite{brahms,brahms-07} and
STAR \cite{star} Collaborations.

The new results of this paper are the following :

\begin{itemize}

\item 

Using 
the simple formula (\ref{120}) adopted from ref.~\cite{knpsj-05}
and
based on Glauber-Gribov multiple
interaction theory and the AGK cutting rules,
we calculated high-$p_T$ hadron production at
large $x_F$ and found a substantial suppression. 
This parameter-free
calculation agrees with recent measurements performed by the BRAHMS
\cite{brahms,brahms-07}
and STAR \cite{star}
Collaborations at forward rapidities in deuteron-gold collisions at RHIC.
Our simple explanation is based on just energy conservation
reflecting a small
survival probability of LRG in multiple quark interactions
at large $x_F$.

\item

With the same input, we explain for the first time very strong
nuclear suppression for $\pi^0$ production at $\eta = 4.0$
measured recently by the STAR Collaboration \cite{star}.
This suppression is more than a factor of $2$ larger 
than at $\eta = 3.2$ (see Fig.~\ref{brahms})
investigated by the BRAHMS Collaboration
\cite{brahms} for $h^-$ production
and no other models with a reasonable alternative
description are known.

\item

In order 
to exclude differences
affected by isospin effects
in production of different species of particles,
we performed also model predictions for
nuclear suppression
at large $p_T$ as a function of pseudorapidity
changing from $3.0$ to $4.0$ (see Fig.~\ref{eta-pi0}).
Predicted a 
huge rise of nuclear suppression 
with $\eta$ 
about a factor of $2$ 
follows from much smaller
survival probability of LRG in multiple quark rescatterings.

\item 

Using proper fragmentation functions from ref.~\cite{fs-07}
we performed model calculations of nuclear suppression
at large $p_T$ and $\eta = 3.0$ for $\pi^+$ and $K^+$
production in a good agreement with the latest data
from the BRAHMS Collaboration \cite{brahms-07}
(see Fig.~\ref{brahms-07}).

\item 

As a consequence of $x_F$ scaling,
we predict approximately the same
nuclear effects at different energies and pseudorapidities
corresponding to the same values of $x_F$ (see
Fig.~\ref{scaling}).
Such a $x_F$ scaling can be verified
in the future by the
BRAHMS or STAR Collaborations mainly at lower energies,
where no effect of coherence is possible. 
It allows to exclude the saturation models or
the models based on CGC from explanation
of nuclear suppression.

\item 

According to $x_F$ scaling we expect
similar nuclear effects also
at midrapidities in the RHIC energy range.
However, the corresponding values of $p_T$ 
for the produced hadrons should be much higher
than at forward rapidities
to keep the same value of $x_F$,
where the nuclear suppression is studied.
In this kinematic region,
investigation of nuclear suppression in production
of different hadrons in $p(d)-Au$ collisions
is also very important because
at large $p_T$
the data cover rather large $x_2\sim 0.05-0.1$
where no effect of coherence is possible.
It gives another possibility to exclude
the models based on CGC.

\end{itemize}

\vspace*{0cm}

\begin{acknowledgments}

This work was supported in part by Fondecyt (Chile) grant 1050519,
by DFG (Germany)  grant PI182/3-1, by the Slovak Funding
Agency, Grant No. 2/7058/27 and by the
grant VZ MSM 6840770039, and LC 07048 (Czech Republic).

\end{acknowledgments}

\end{document}